# Probing Optical Vortex Beams via a Controllable Anisotropic Diffractive Phase Element

Ali Mardan Dezfouli,[1*] Mario Rakić,[1] Hrvoje Skenderović,[1,2]

[1]*Centre for Advanced Laser Techniques, Institute of Physics, 10000 Zagreb, Croatia,*
[2]*Centre of Excellence for Advanced Materials and Sensing Devices, Photonics and Quantum Optics Unit, Institute Ruđer Bošković, Zagreb 10000, Croatia*
*Corresponding author: amdezfouli@ifs.hr*

**Abstract:**

In this work, we investigate the diffraction of optical vortex beams through a tunable elliptical Fresnel phase mask (TEFPM). The resulting diffraction patterns are influenced by both the topological charge of the beam and the ellipticity of the mask, exhibiting characteristic intensity distributions that allow direct determination of the magnitude and sign of the topological charge. Requiring only a single controllable parameter, the TEFPM offers a simple and adaptable approach for vortex charge characterization, with a reduced detection distance that supports optical system compactness. The experimental results are confirmed by the exact analytical solutions.

______________________________________________________________________

It is widely recognized that light can carry orbital angular momentum (OAM). The OAM is directly linked to the helical phase structure of the optical vortex [1]. This characteristic spiral phase results in singularity along the propagation axis, which manifests as a dark core in the beam intensity profile. The phase structure of vortex beams is described by $exp(il\phi)$, where $\phi$ is the azimuthal angle, and $l$ represents the twisting number widely known as topological charge (TC). Each photon in such a beam carries an OAM in the direction of propagation. The topological charge of an optical vortex beam represents the number of $2\pi$ phase twists around its axis and can be either positive or negative.

In recent years, the study of vortex beam has attracted significant interest due to its broad range of applications, particularly in areas like high capacity optical communication [2,3,4], quantum information processing [5,6], optical tweezers [7], and more recently the high capacity OAM multiplexing holography [8], remote sensing [9] and nano-plasmonic [10].

Determining the topological charge value is a task of critical importance in numerous applications, and several approaches have been developed to achieve this. Certain techniques [11–14] determine TC through interferometry, depending on a complicated setup. These approaches necessitate the use of two separate beams to extract the TC information.

A class of static methods uses carefully designed apertures, such as triangular, single slit, double-slit, and annular shapes [15-18], to study vortex beams diffraction without need for extra beam. However, these methods rely on precise central illumination; misalignment significantly degrades diffraction patterns and causes energy loss, reducing efficiency. Another set of static approaches employs passive optical elements like cylindrical lenses [19], tilted convex lenses [20], mode convertor [21] and off-axis parabolic mirrors [22,23]. These methods are limited by their fixed geometries and lack of tunability, making them inflexible for different experimental conditions. Their bulky designs further hinder integration into compact optical systems, posing a significant challenge for applications requiring space efficiency. In addition, inherent aberrations and manufacturer fabrication imperfections can distort the results, reducing the accuracy and reliability of topological charge detection.

Another class, dynamic in nature, is diffraction gratings, which became popular for TC characterization due to their superior adaptability and tunable capabilities compared to aperture-based methods. However, this often comes with increased complexity, as many require tuning multiple parameters, complicating the design and fabrication. For example, parabolic-line linear gratings, it requires five different parameters [24] sinusoidal phase gratings and annular radial gratings with varying periods each need three parameters [25,26], gradually-changing-period grating [27] require two, while single-parameter annular grating [28] is simpler but highly sensitive to beam position which makes it less robust. Although gratings-based methods offer good performance, relying on accurate parameter tuning and alignment sensitivity in some cases limits their practical ease and applicability, motivating simpler yet robust alternatives.

In this paper, we propose a simple method for detecting the topological charge of optical vortex beams using a tunable elliptical Fresnel phase mask (TEFPM). Unlike conventional diffraction-based techniques that often require complex alignment or multiple tunable parameters, our approach exploits the anisotropic diffraction characteristics of the TEFPM for straightforward TC characterization at focal plane. The simplicity of the approach is reflected in the TEFPM design, which relies on only one tunable parameter, the ellipticity. This minimal parameterization, offers a practical balance between robustness and design simplicity. Compared to our previous approach using a 101.6 mm fixed focal length of parabolic mirror, the TEFPM reduces the detection distance to 50 mm, offering a shorter propagation distance that directly supports system compactness. We provide an analytical description of the TEFPM diffraction behavior by vortex and validate it experimentally. The results demonstrate good agreement between theory and measurement. Fresnel zone plates, on which the TEFPM design is based, are widely used diffractive elements, valued for their simple geometry and effective focusing properties. Their applications span from high-resolution imaging at the atomic scale and X-ray microscopy [29,30], to wavefront control in structured light [31] and more recently to antenna technology for nano-satellites [32].

The complex amplitude of the optical vortex in Cartesian coordinate can be written [33]:

$$E(x_0, y_0) = (x_0 \pm i y_0)^{|l|} \exp\left(-\frac{x_0^2 + y_0^2}{w_0^2}\right) \tag{1}$$

where $l$ and $w_0$ indicate the topological charge and radius parameter of the beam at $z = 0$, respectively. Moreover, $+i$ or $-i$ is used when value of $l$ is positive or negative, respectively. Prior research has primarily addressed circular Fresnel zone plate [33], in the present work we consider more general case by considering ellipticity as the main tunable parameter, enabling controlled modification of the vortex beam diffraction and direct determination of its topological charge

The complex-amplitude transmittance of the TEFPM pupil can be expressed as a Fourier series:

$$T(x_0, y_0) = \sum_{m=-\infty}^{\infty} t_m \exp\left(\frac{im\pi}{\lambda f}\left(\frac{x_0^2}{a^2} + \frac{y_0^2}{b^2}\right)\right) \tag{2}$$

The coefficients $t_m$ represent the expansion terms, and $f$ is the first focal length for an incident wave of wavelength $\lambda$, when the TEFPM functions as a diffractive lens. After the beam passes through an aperture defined by a transmission function $T(x, y)$, the complex amplitude of the light immediately beyond the aperture becomes $E_T(x_0, y_0) = T(x_0, y_0) E(x_0, y_0)$. For a coherently illuminated pupil, described by a complex-amplitude transmittance function, the diffracted field distribution can be obtained from the Fresnel diffraction integral [34]:

$$E(x_i, y_i) = \frac{e^{ikz}}{i\lambda z} e^{i\alpha(x_i^2 + y_i^2)} \iint_{-\infty}^{+\infty} E_T(x_0, y_0) \exp[i\alpha(x_0^2 + y_0^2)] \exp[-2i\alpha(x_i x_0 + y_i y_0)] dx_0 dy_0 \tag{3}$$

To obtain a closed-form solution, we used tabulated integrals from [35]. The resulting expression is given by:

$$E(x_i, y_i) = C \sum_{m=-\infty}^{\infty} t_m \sum_{k=0}^{|l|} \binom{|l|}{k} (\pm i)^{|l|-k} \left[ \gamma_x^{-\left(\frac{k+1}{2}\right)} \gamma_y^{-\left(\frac{|l|-k+1}{2}\right)} \exp\left(\frac{\beta_x^2}{4\gamma_x}\right.\right.$$
$$\left.\left. + \frac{\beta_y^2}{4\gamma_y}\right) H_k\left(\frac{i\beta_x}{2\sqrt{\gamma_x}}\right) H_{|l|-k}\left(\frac{i\beta_y}{2\sqrt{\gamma_y}}\right)\right]$$
(4)

Where

$$\gamma_x = -\frac{1}{w_0^2} + \frac{im\pi}{\lambda f a^2} + i\alpha, \quad \gamma_y = -\frac{1}{w_0^2} + \frac{im\pi}{\lambda f b^2} + i\alpha,$$

$$\beta_x = 2i\alpha x_i, \beta_y = 2i\alpha y_i, \quad C = \frac{\pi e^{ikz}}{i\lambda z} \left(\frac{i}{2}\right)^{|l|} e^{i\alpha(x_i^2 + y_i^2)}, \quad \alpha = \frac{k}{z}, \quad k = \frac{2\pi}{\lambda}$$

(5)

Where in equation (4), The solution contains Hermite polynomials, with orders $k$ and $|l| - k$ for the $x_i$ and $y_i$ directions respectively. This result show that the diffraction pattern distort due to different phase curvature in $x_i$ and $y_i$ directions while the spatial distribution becomes elliptical due to quadratic phase terms of TEFPM and parameters $a$ and $b$.

Figure 1 shows the theoretical results for intensity and phase distributions of optical vortex ($l = 5$) diffracted by the TEFPM with $f = 50mm$. The results are shown for three different ellipticity parameters for (a) $a = 1, b = 1$ (circular case), (b) $a = 1.2, b = 1$, and (c) $a = 1.5, b = 1$, where $a$ and $b$ denote the scaling factors along the $x$- and $y$-directions, respectively. The schematic at top of Fig.1. additionally shows the geometrical focus point of the TEFPM.

In the circular configuration ($a = 1, b = 1$), the vortex beam is focused at the designed focal plane in a symmetric manner. The diffraction pattern maintains its circular symmetry and the vortex is only compressed to a smaller scale, without introducing any distortions. Both the intensity and phase distributions remain consistent with the undistorted helical structure of the vortex beam. When ellipticity is introduced by increasing $a$ while keeping $b = 1$, the situation changes significantly. As results show the diffraction pattern at the focal plane is no longer circularly symmetric, instead, the vortex field becomes elongated and distorted while having fringes. When the mask is elliptical ($a \neq b$), this introduces an astigmatic like phase modulation, causing the vortex to elongated at slightly different scales along the two axes, producing a distorted diffraction pattern at the proximity to focal plane.

Under these conditions, the topological charge of the vortex can be identified directly from the diffraction pattern: the number of dark (or equivalently bright) fringes corresponds to the magnitude of the charge ($|l| + 1$ bright spots or $|l|$ dark lines ), while the orientation of these fringes distinguishes the sign of vortex beam. As results show, for the positive topological charges, the fringes align along a diagonal direction, whereas for negative charges, the orientation rotates to an anti-diagonal.

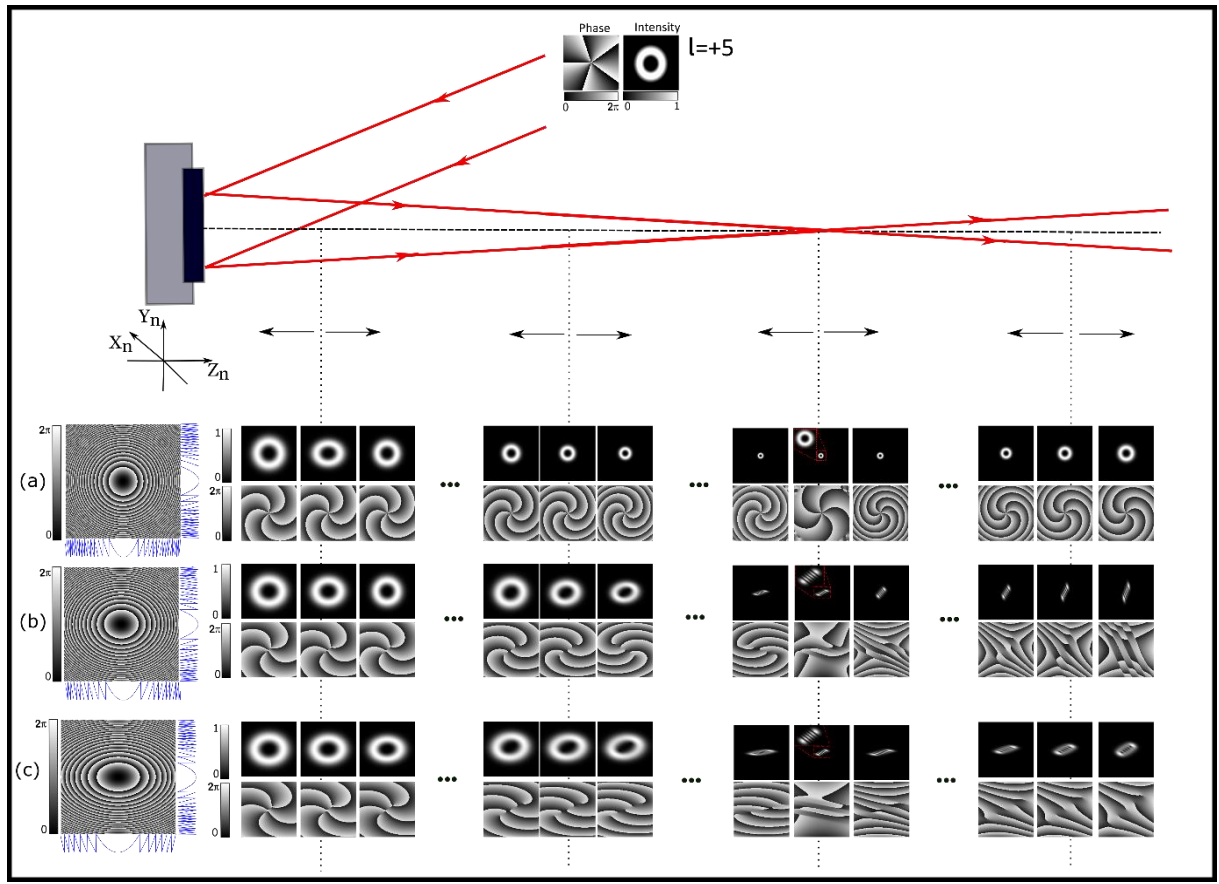

Figure 1: Theoretical results for intensity and phase distributions of optical vortex beams diffracted by the TEFPM at the focal plane. The patterns correspond to different ellipticity parameters: (a) $a = 1, b = 1$, (b) $a = 1.2, b = 1$, (c) $a = 1.5, b = 1$, where $a$ and $b$ are scaling factors along $x$ and $y$ directions, respectively. The parameters are vortex beam has $TC = +5$ and focal length $f = 50mm$, beam waist $0.3\ mm$, and $\lambda = 632.8nm$. The profile of the TEFPM is illustrated for each respective mask in horizontal and vertical direction (see visualization V1-V3).

Fig.2. shows the theoretical results for intensity and phase distributions of optical vortex beams diffracted by the TEFPM with $f = 100mm$ for the same parameters as in Fig.1.

Regarding the phase, theoretical results show that the helical phase of the vortex beam appears to reverse its rotation when propagating through the focal point of the TEFPM. This behavior is consistent with previous observations of Gouy phase effects in optical vortices [36]. The apparent rotation reversal does not indicate a change in topological charge, which is conserved under free-space propagation. Rather, it arises from the axial phase evolution near the focus which is the combination of the Gouy phase shift and the inversion of wavefront curvature across the focal plane effectively transforms the azimuthal coordinate. As a result, the transverse phase pattern appears to rotate in the opposite direction after the focal point. For a clearer understanding of the of this and the impact of ellipticity on the diffracted field, we include supplementary (see Visualization V1-V3 and V4-V6) for phase and intensity, corresponding to different ellipticity settings for Fig.1 and Fig.2, respectively.

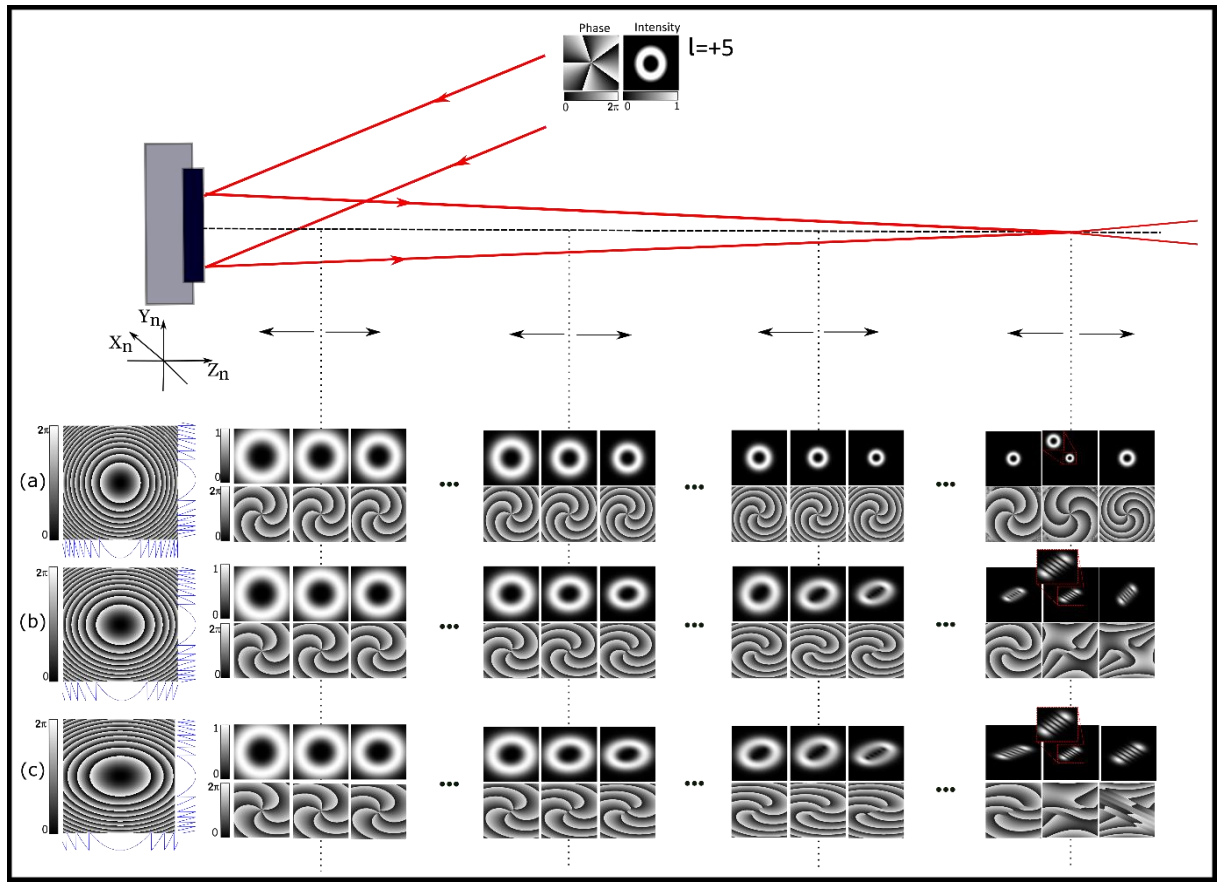

*Figure 2: Theoretical results for intensity and phase distributions of optical vortex beams diffracted by the TEFPM with $f = 100mm$. The patterns correspond to phase mask with different ellipticity parameters: (a) $a = 1, b = 1$, (b) $a = 1.2, b = 1$, (c) $a = 1.5, b = 1$, where a and b are scaling factors along x and y directions, respectively. The parameters are vortex beam has $TC = +5$ and focal length $f = 100mm$, beam waist $0.3\ mm$, and $\lambda = 632.8nm$. The profile of the TEFPM is illustrated for each respective mask in horizontal and vertical direction while the schematic at top additionally indicates the geometrical focus point of the TEFPM, (see visualization V4-V6).*

To have better understanding of the beam dynamics, Fig. 3 shows the vortex wave field ($l = 5$) 3D propagation in the X-Z and Y–Z plane after TEFPM, obtained from theoretical results of the diffraction. The 3D view reveals how the ellipticity of the TEFPM influences the axial evolution of the vortex beam, leading to characteristic distortions in the diffraction pattern. The analysis shows the axial elongation and asymmetric redistribution of intensity along the propagation axis. Subplots ($a1$–$a3$) correspond to a focal length of $50\ mm$, while ($b1$–$b3$) correspond to $100\ mm$, show a clear comparison of propagation behavior at different focusing conditions for different ellipticity parameters.

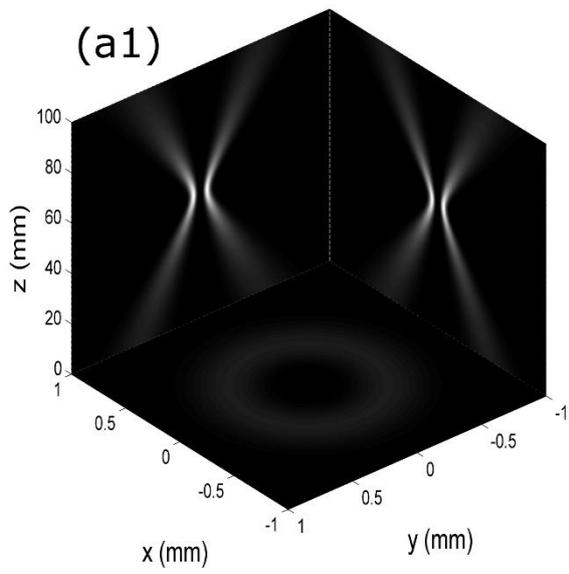 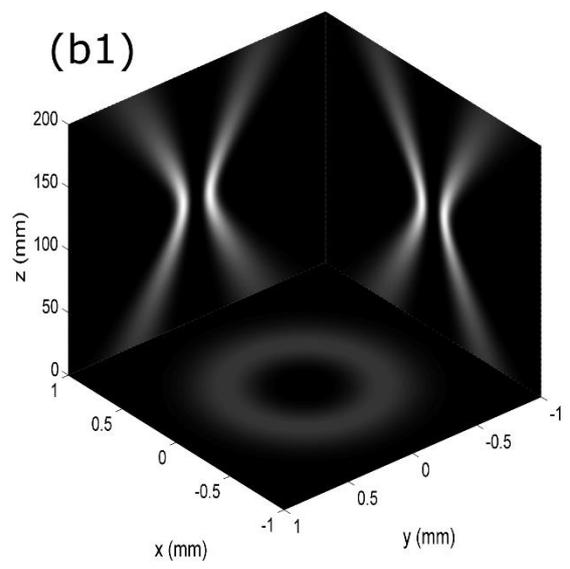

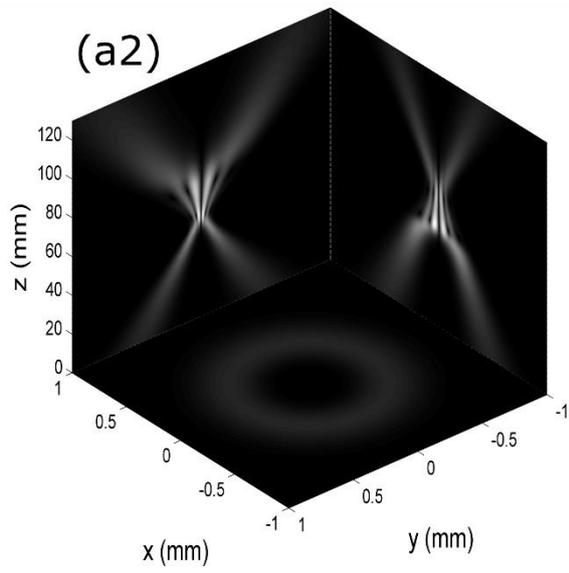 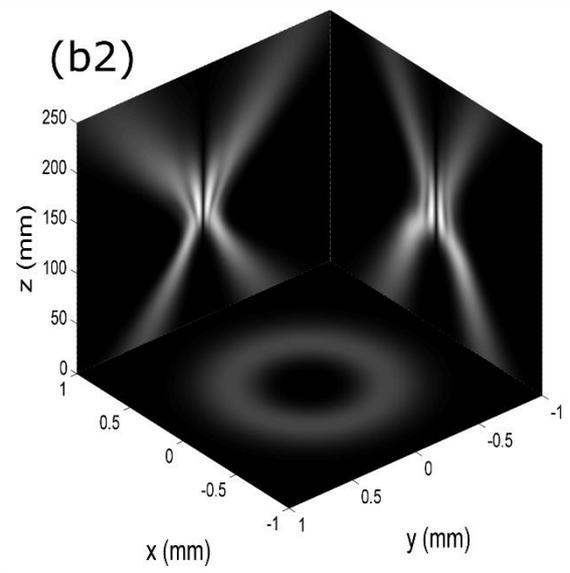

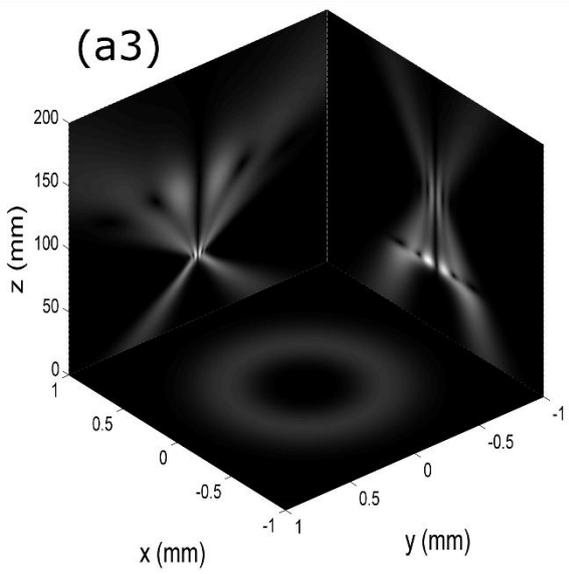 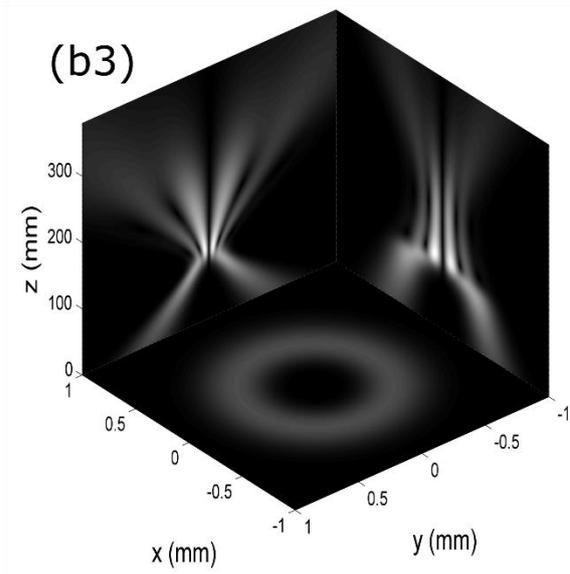

*Figure 3: 3D Propagation of a vortex beam (l=5) through the TEFPM in X–Z and Y–Z planes, illustrating diffraction dynamics for f=50 mm (a1-a3) and f=100 mm (b1-b3) under different ellipiticity condition. First rows correspond to ($a = 1, b = 1$), second row ($a = 1.2, b = 1$), and third row ($a = 1.5, b = 1$).*

To confirm our theoretical predictions, we built an experimental setup to characterize the topological charge of vortex beams using an TEFPM as shown in Figure 4. We used a He-Ne laser with a wavelength of 632.8 nm and an output power of about 21 mW, producing a fundamental Gaussian beam with a spot size of 0.8 mm. A neutral density filter (NDF) was placed to reduce the beam intensity to avoid saturating the camera. A spatial filtering system (SF) was used to clean the beam. Vortex beams were created by displaying computer-generated holograms on the first SLM as SLM1, which applied a blazed modulation to encode the desired vortex phase pattern. To select only the first-order diffracted beam, a pinhole in the Fourier plane blocked unwanted diffraction orders. This vortex beam then illuminated the second SLM shown as SLM2 which is loaded by TEFPM. The setup utilized two phase-only spatial light modulators (Holoeye, 1920 × 1080 resolution). In the experiment, a linear phase ramp was added to the TEFPM to spatially separate the desired diffraction order from the undiffracted zero order and back reflection. A CCD camera recorded the diffraction pattern produced by the TEFPM.

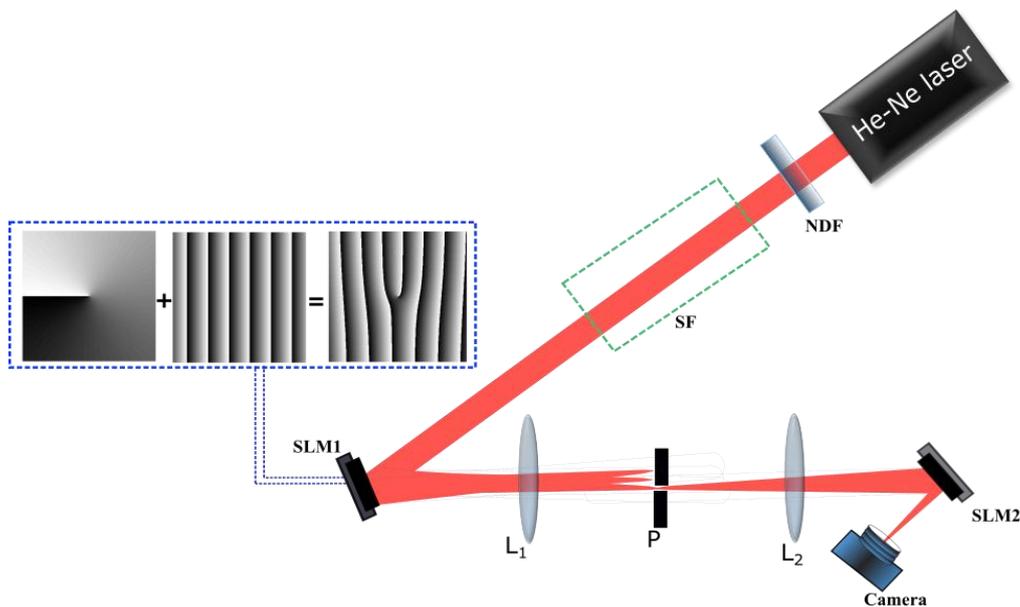

*Figure 4: The schematic depicts the experimental setup. SF: spatial filtering, L1, L2: Lens, NDF: neutral density filter, SLM1 and SLM2: spatial light modulator.*

Figure 5 presents the experimental diffraction patterns of optical vortex beams diffracted by the TEFPM. The experimental results of vortex beam focusing by the TEFPM are shown for two focal lengths $f = 50mm$ and $f = 100mm$. For each case, the left panels correspond to an ellipticity of $a = 1.2, b = 1$ while the right panels correspond to $a = 1.5, b = 1$.

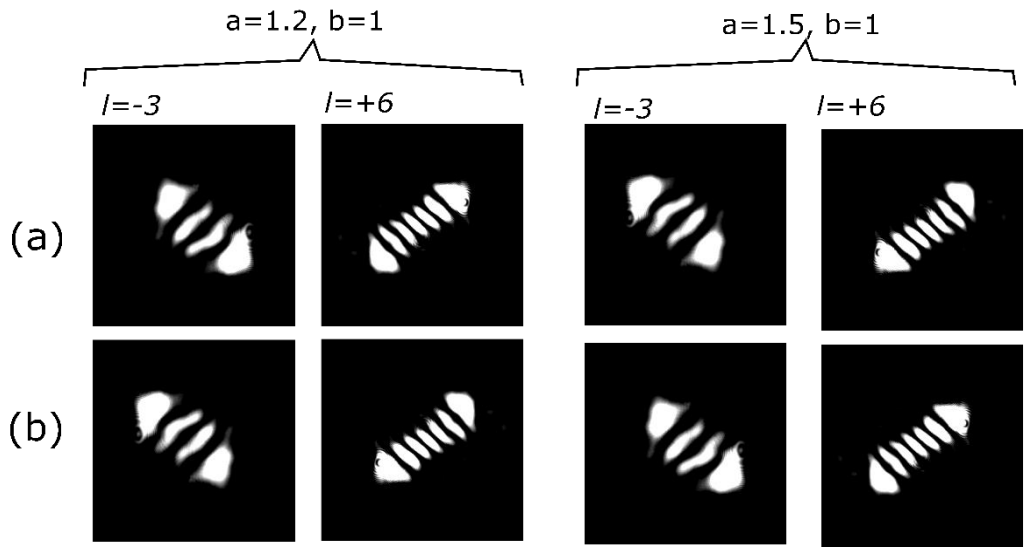

Figure 5: Experimental results for vortex focusing by the TEFPM. Row (a) corresponds to f=50mm and row (b) to f=100mm. The left block shows the case for ellipticity $a = 1.2, b = 1$ while the right block shows results for ellipticity $a = 1.5, b = 1$.

For focal lengths ($f = 50mm$ in row (a) and $f = 100mm$ in row(b) in figure 5), we capture diffraction patterns at proximity to focal length. The recorded intensity distributions show good agreement with the theoretical predictions, both in the overall structure and in the characteristic distortions introduced by ellipticity.

In summary, we employed a tunable elliptical Fresnel phase mask to extract both the magnitude and sign of TC from vortex beams. Our results demonstrate that the diffraction patterns generated by TEFPM provide a reliable and straightforward means for TC detection, with the number of dark fringe cores corresponding to the absolute value of TC and the orientation of the intensity distribution revealing its sign.

In contrast to other grating-based methods, which often require simultaneous control of multiple parameters for detection, the TEFPM offers a streamlined solution. Its operation relies on tuning only a single parameter, the ellipticity, while still providing flexibility in focal length adjustment. Compared to other static approach using optical elements, which are inherently passive and bulky and susceptive to manufactory imperfections, the TEFPM achieves compactness and adaptability without compromising performance. Our experiments demonstrated well-defined fringes at the effective focal plane, with strong visibility near the focal region.

An important advantage of our approach lies in its potential for its scalability and space efficiency. Compared to previous work, using a parabolic off-axis mirror with a fixed focal length of 101.6 mm, which required extended detection distances. By contrast, the TEFPM enables probing topological charge with focal lengths as short as 50 mm, effectively reducing the required propagation distance for TC detection. This dynamic reduction in detection distance is particularly valuable in scenarios where compactness is critical in integrated optical systems design.

Further our results reveal that the diffraction of vortex beams by the TEFPM exhibits a rich and dynamic evolution, which we have shown can be directly harnessed for determining both the magnitude and sign of the TC. Beyond serving as a practical detection method, these results reveal that TEFPM can be itself be a promising avenue for exploring new aspects of structured light diffraction. Looking ahead, our future efforts will extend this approach toward broader regimes, examining other beam types carrying OAM, higher-order Laguerre Gaussian (LG) beam, and TEFPM fabrication using lithography-based methods for micro design, to further expand the applicability of this method in optical metrology and OAM-based applications.

**Funding** Croatian Science Foundation (DOK-2020-01-8574), Ministry of Science and Education of Republic of Croatia grant No. KK.01.1.1.01.0001, and the project Fab - fs financed by the European Union through the National Recovery and Resilience Plan 2021-2026 (NRPP).

**Disclosures.** The authors declare no conflicts of interest.

**Data Availability**

The data that support the findings of this study are available from the corresponding author upon reasonable request.